\documentclass[letterpaper,aps,twocolumn,preprintnumbers]{revtex4}
\usepackage{graphicx}
\begin{document}

\title{Semiclassical Quantization Rule for Bound-State Spectrum in Quantum Dots: Scattering Phase Approximation}
\author{Wei Chen}
\affiliation{Department of Physics, 
National Tsing-Hua University,
Hsinchu 300, Taiwan, R.O.C.}
\author{Tzay-Ming Hong}
\affiliation{Department of Physics, University of Texas, Austin, TX 78712, U.S.A.}
\affiliation{Department of Physics, 
National Tsing-Hua University,
Hsinchu 300, Taiwan, R.O.C.}
\author{Hsiu-Hau Lin}
\affiliation{Department of Physics, 
National Tsing-Hua University,
Hsinchu 300, Taiwan, R.O.C.}
\affiliation{Physics Division, National Center for Theoretical Sciences, Hsinchu 300, Taiwan, R.O.C.}
\date{\today}

\begin{abstract}
We study the quantum propagator in the semiclassical limit
with sharp confining potentials. Including the energy-dependent scattering phase due to sharp confining potential, the modified Van Vleck's formula is derived. 
We also discuss the close relations among quantum
statistics, discrete gauge symmetry, and hard-wall
constraints.  Most of all, we formulate a new quantization
rule that applies to {\it both} smooth and sharp boundary
potentials.  It provides an easy way to compute quantized
energies in the semiclassical limit and is extremely useful
for many physical systems.
\end{abstract}

\maketitle

\section{Introduction}

The most straightforward method to obtain the bound-state spectrum for a quantum system is to solve the Schr\"odinger equation. However, if the potential profile is smooth compared with the wavelength of  the particle, the energy spectrum can be obtained by the semiclassical WKB approximation. The semiclassical approach reduces the task of solving the differential equation into a simple integral. While the great simplification is attractive, it does not work all the times. In quantum dots, the confining potential is usually sharp and leads to strong quantum interferences which invalidate the semiclassical approach. One notices that the semiclassical approximation only breaks down near the sharp confining potential. It motivates us to generalize the conventional semiclassical approach by including the quantum interferences exactly near the turning points where the semiclassical approximation is not appropriate. Rather nicely, we were able to capture the complicated quantum interference effects by a simple energy-dependent scattering phase correction. 

To elucidate this point, it is convenient to adapt the path integral formulism.  
Path integral provides an alternative approach to formulate quantum
mechanics\cite{Feynman65,Kleinert95}.
The quantum propagator $G(x,x^{\prime };T)$, that is the key
quantity in quantum mechanics, is shown to equal the
summation over all possible paths with the same end points.
In the semiclassical limit, the dominant contribution comes
from classical trajectories and fluctuations around 
them\cite{Sepulveda92,Manning96,Stock97}.
Within the stationary phase approximation including
fluctuations up to the quadratic order, the quantum
propagator can be approximated by the Van Vleck's
formula\cite{VanVleck28}.
In general, there would be many classical trajectories that
satisfies the same boundary conditions, and the Berry phase
interferences between them are
important\cite{Berry72,Gutzwiller90,Gutzwiller71}.
By Morse's theorem, the second variation, considered as
quadratic fluctuations around a given trajectory from
$x^{\prime }$ to $x$ in time $T$, has as many negative
eigenvalues as there are conjugate (turning) points along
the trajectory.
These conjugate points give rise to a Berry phase $\nu \pi
/2$ for the trajectory, where $\nu $ is the total number of
conjugate points along the trajectory, or sometimes referred
to as the Maslov or Morse index\cite{Maslov81}.

Not only elucidating the crossover between classical and
quantum mechanics, the semiclassical limit also provides a
convenient way to calculate the bound state energy.  Instead
of solving the Schr\"{o}dinger equation directly, the bound
state spectrum can also be computed by the WKB
approximation\cite{Sakurai94}. 
In order to account for the interference effects among
classical trajectories correctly, we rederive Van Vleck's
formula with an extra scattering phase correction due to
sharp confining potentials. 

Following the standard stationary phase approximation and
making a Legendre transformation of the time variable in the
quantum propagator to the energy variable, we are able to
generalize the Einstein-Brillouin-Keller (EBK) quantization
rule\cite{Einstein17,Brillouin26,Keller58} 
with an additional phase correction term
\begin{equation}
\oint \sqrt{2m[E-V(x)]}dx=2n\pi +\sum_{s} \phi_{s}(E),  \label{NewEBK}
\end{equation}
where $\phi_{s}(E)$ is the energy-dependent scattering phase due to collisions with the confining potential. The usual WKB approximation is the special case where the scattering phase at each turning point is assume to take on the energy-independent value $\phi_{s}(E) = \pi/2$. On the other hand, if the confining potential becomes infinitely sharp (hard-wall limit), the scattering phase rises to $\pi$. The modified EBK quantization rule in Eq.$~$(\ref{NewEBK})
relaxes the requirement of the potential smoothness in the
WKB approximation. This is of great advantage because many physical systems
including quantum dots, quantum wells, Hall bars, electronic
wave guides, etc., have both hard-wall-like potentials (from
sample edges) as well as smooth potentials (by applying
external fields) at the same time.

The paper is organized in the following way. 
In Sec.  II, we introduce the Van Vleck's formula and apply
it to simple systems. We explicitly show that the Van Vleck's approximation is
incorrect in the presence of hard walls and the scattering phase correction is crucial.
In Sec III, we compute the energy dependence of the scattering phase at each turning point and derive the modified Van Vleck's formula.
In Sec. IV, we derive the key result of this paper --
the modified EBK quantization rule.
We apply it to physical systems with both smooth and hard
confinement potentials and show that the modified term is
necessary to obtain the correct energy levels. 
Finally, in Sec V, we relate the connection between the quantum statistics, discrete gauge symmetry to the scattering phase approach.
Then a brief conclusion follows.

\section{Quantum Propagator and Classical Trajectories}

In the path integral formalism\cite{Feynman65}, the quantum
propagator equals the sum over all possible paths with the
same end points,
\begin{eqnarray}
G(x,x^{\prime};T) &\equiv& \langle x| e^{-iHT}
|x^{\prime}\rangle \nonumber
\\
&=& \int_{x^{\prime}}^{x} {\cal D}[x] \exp \left( i
\int_{0}^{T} L(x,\dot x, t) dt \right),
\end{eqnarray}
where the measure ${\cal D}[x]$ denotes all possible paths
with end points $x(0)=x^{\prime}$ and $x(T)=x$.
In the semiclassical limit, the phase inside the path
integral oscillates rapidly except in the neighborhood of
the classical trajectories.
Within the stationary phase approximation including
fluctuations up to quadratic order, the propagator is
approximated by the Van Vleck's formula,
\begin{equation}
G(x,x^{\prime};T) \simeq \frac{1}{\sqrt{2\pi i}}\sum_{p} \sqrt{C_{p}} 
\exp[iA_{p} -i\nu_{p} \frac{\pi}{2} ],  \label{VanVleck}
\end{equation}
where $A_{p}(x,x^{\prime};T)$ is the action of the classical
trajectory starting from $x(0)=x^{\prime}$ and ending at
$x(T)=x$, and the subscript $p$ denotes all classical paths
with the desired end points.  The strength of the quadratic
fluctuations\cite{Gutzwiller90} around the classical
trajectory is
\begin{equation}
C_{p}= \left| - \frac{\partial^{2} A}{\partial x \partial x^{\prime}}\right|.
\end{equation}
Finally, the total number of conjugate (or turning) points
along the classical trajectory is denoted by $\nu$.  Notice
that, for each conjugate point, there is a $\pi/2$ Berry
phase associated with it.  Van Vleck's formula provides a
completely classical approximation of the quantum
propagator, in the sense that all relevant elements can be
computed from the classical trajectories.

A straightforward example of the Van Vleck's formula is a
free particle moving on the a finite ring with length $L$. 
There are infinite classical paths which satisfy the
conditions $x(0)=x^{\prime }$ and $x(T)=x$.  The total
(route) distance of each classical trajectory is
$d_n=x-x^{\prime }+nL$, where $n$ is an integer.  The action
for each trajectory is
\begin{equation}
A_n(x,x^{\prime };T)=\frac m{2T}(x-x^{\prime }+nL)^2.
\end{equation}
Taking the derivative of the action, the strength of
fluctuations around each trajectory $C_n=m/T$ is independent
of the end points and the choice of trajectories.  Since the
particle moves at constant velocity, it is obvious that
there is no conjugate point along any classical trajectory
and thus $\nu _n=0$.  Besides, because the fluctuations of
the classical trajectory of a free particle are exactly
quadratic, we expect the Van Vleck's formula to be exact for
this system,
\begin{equation}
G(x,x;T)=\sqrt{\frac m{2\pi iT}}\sum_n\exp 
\left[ i\frac{m}{2T}(x-x^{\prime }+nL)^2\right]
\end{equation}
This infinite sum can be re-written in terms of its Fourier
function with the use of Poisson summation formula in
Appendix A.\cite{Book} Notice that
\begin{equation}
f(y)=e^{i\alpha (y+\beta )^2} \leftrightarrow 
F(p)=\sqrt{\frac{i\pi }\alpha }
e^{-ik^2/4\alpha +ik\beta }
\end{equation}
Choosing $a=L$, the summation over coordinate $y=na$ can be
turned into the summation over momentum $k_n=2n\pi /L$.  The
propagator is then
\begin{eqnarray}
G(x,x;T)=\frac 1L\sum_n\exp [ik_n(x-x^{\prime })-iE_nT],
\end{eqnarray}
where $k_n=2n\pi /L$ is the quantized momentum and
$E_n=k_n^2/2m$ is the quantized energy.  It is obvious that
the propagator $G(x,x^{\prime };T)$ calculated by the Van
Vleck's formula is exact in this case.

Let us now apply the Van Vleck's formula to another physical
system -- a free particle bouncing back and forth between
two hard walls.  We calculate the propagator
explicitly and show that the Van Vleck's formula leads to
incorrect results.

\begin{figure}[tbp]
\centering\includegraphics[width=6cm]{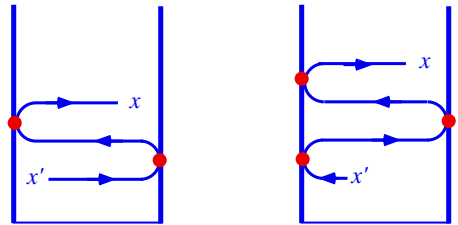}
\caption{Classical trajectories in the presence of two hard walls. On the
left is a trajectory with even reflection points $r=2,$ while the right with
odd reflection points $r=3$.}
\end{figure}

The trajectories in this problem can be classified by the
number of collisions with the hard walls, as seen in Fig. 
1.  For those trajectories that collide with the hard walls
even times, the route distance is $ d_n^e=x-x^{\prime
}+2nL$, while the distance is $d_n^o=x+x^{\prime }+2nL$ for
trajectories that collide with the walls odd times.  The
action for each trajectory can be computed straightforwardly
\begin{eqnarray}
A_n^e(x,x^{\prime };T) &=&\frac m{2T}(x-x^{\prime }+2nL)^2, \\
A_n^o(x,x^{\prime };T) &=&\frac m{2T}(x+x^{\prime }+2nL)^2.
\end{eqnarray}
Here $A^{e/o}(x,x^{\prime };T)$ denotes the action for
trajectories with even/odd reflection points.  The
fluctuations along all trajectories contribute the same
$C_n=m/T$ as in the previous example.  For an
one-dimensional motion, a conjugate point is identified as
the position where the velocity vanishes.  However, for a
free particle bouncing back and forth between two hard
walls, the velocity is constant up to a minus sign and does
not vanish at any point along the classical trajectory. 
Thus, the number of conjugate points is zero, $\nu =0$.

The propagator without any phase correction is
\begin{eqnarray}
G_{VV}(x,x^{\prime };T) &=&\sqrt{\frac m{2\pi
iT}}\sum_n\bigg\{\exp [i\frac m{2T}(x-x^{\prime }+2nL)^2]
\nonumber \\
&+&\exp [i\frac m{2T}(x+x^{\prime }+2nL)^2]\bigg\}.
\end{eqnarray}
Both infinite sums can be turned into summations over
discrete momentum again by mean of the Poisson summation
formula.  The prefactors cancel as in the previous example
and we are left with the simple result,
\begin{eqnarray}
\lefteqn{G_{VV}(x,x' ;T)=\frac 1L\sum_{n=0}^\infty \exp [-iE_nt]\times }
\nonumber \\
&\times &\bigg\{ \cos [k_n(x-x^{\prime })]+\cos [k_n(x+x^{\prime })]\bigg\},
\label{Mistake}
\end{eqnarray}
where $k_n=n\pi /L$ is the quantized momentum and
$E_n=k_n^2/2m$ is the quantized energy.  Combining two
cosines would leads to $\cos (k_nx)\cos (k_nx^{\prime }),$
while the correct form should be $\sin (k_nx)\sin
(k_nx^{\prime })$.  In fact, one can recover the exact answer
(with all prefactors right!)  if we change the sign of the
second term in Eq.~(\ref{Mistake}).  That is, only if we
assign an extra phase $\pi $ to trajectories with {\it
odd} reflection points, will the modified Van Vleck's
formula become correct!

In the following section, we study the path integral
formalism in the presence of a single hard-wall boundary and
show that an extra phase correction arises naturally due to collisions with the confining potential.

\section{Scattering Phase Due to Hard Wall}

Consider a particle moving under the influence of a regular
potential $V(x)$ and a hard-wall potential $V_c(x)$.  The
Hamiltonian is
\begin{equation}
H=\frac{p^2}{2m}+V(x)+V_c(x),
\end{equation}
where $V_c(x)$ is the hard-wall potential at $x=0$, 
\begin{eqnarray}
V_c(x)=\left\{ 
\begin{array}{cl}
0, & x>0; \\ 
\infty , & x<0.
\end{array}
\right. 
\end{eqnarray}
The regular potential is treated in the ordinary way while
the hard-wall one is viewed as the depletion of Hilbert
space.  The complete set of the Hilbert space is now
reduced,
\begin{eqnarray}
\int_0^\infty dr|r\rangle \langle r| &=&{\bf 1}, \\
\sum_{\phi =0,\pi }\int \frac{dp}{2\pi }e^{i\phi }|p\rangle
\langle e^{i\phi }p| &=&{\bf 1}.
\label{Bases}
\end{eqnarray}
It would become clear later that the phase $\phi $ is
associated with the scattering phase in the path integral. 
Slicing the time interval $T$ into infinitesimal pieces and
inserting complete sets of the coordinate space, the
propagator is
\begin{eqnarray}
G(r,r^{\prime };T) &=&\langle r|e^{-iHT}|r^{\prime }\rangle
\nonumber \\
&=&\int_0^\infty dr_n\prod_{n=0}^{N-1}\langle
r_{n+1}|e^{-i\epsilon H}|r_n\rangle ,
\end{eqnarray}
where $r_N=r$ and $r_0=r^{\prime }$ are all positive.  Each
matrix element in the product is computed by inserting the
complete set in momentum space into Eq.(\ref{Bases}),
\begin{eqnarray}
\langle r_{n+1}| &e^{-i\epsilon H}&|r_n\rangle =\int
\frac{dp_n}{2\pi }\exp \left[ -i\epsilon H_n\right]
\nonumber \\
&\times &\sum_{x_n=\pm r_n}e^{ip_n(r_{n+1}-x_n)-i\phi _n},
\end{eqnarray}
where the phase $\phi =0$ for $x_n=r_n$, and $\phi =\pi $
when $x_n=-r_n$.  Since $x_n=\pm r_n$, the two terms can be
combined and lead to the unconstraint integral over $x_n$. 
After changing the constrained variable $ r_n$ to $x_n$, it
is convenient to write the Berry phase $\phi _n$ in the
following way
\begin{equation}
\phi _n=\pi [\Theta (x_{n+1})-\Theta (x_n)].
\end{equation}
Notice that the Berry phase is zero if the path does not
pass through $x=0$ in the infinitesimal time interval $dt_n$
and $\pi $ if the path passes through.  The integral over
momentum can be carried out easily and the propagator is
\begin{equation}
G(r,r^{\prime };T)=\sum_{x^{\prime }=\pm r^{\prime }}e^{i\phi
_s}\int_{x^{\prime }}^r{\cal D}[x]\exp [i{\cal A}(r,x^{\prime };T)].
\label{Propagator}
\end{equation}
The total phase $\phi _s=\pi [\Theta (r)-\Theta
(x^{\prime })]$ is a boundary term and can be pulled out of
the path integral\cite{Kleinert95}.  The paths are divided
into two topologically distinct classes.  For all possible
paths starting from $r$ to $r^{\prime }$, the scattering phase is
zero, while for those starting from $r$ to $-r^{\prime }$,
the scattering phase is $\pi $ that causes a minus sign.  

The classical trajectories among the paths can be then
classified in the same way.  Furthermore, trajectories with
end points $r$ and $r^{\prime }$ can be identified as
trajectories (in the physical half plane) with even
reflection points and those with end points $r$ and
$-r^{\prime }$ are trajectories with odd reflection points.
Therefore, in the semiclassical limit, the Van Vleck's
formula is modified with an extra phase term,
\begin{equation}
G(r,r^{\prime};T) \simeq \frac{1}{\sqrt{2\pi i}}\sum_{p} \sqrt{C_{p}} \exp[i
A_{p} - i\phi_{p}],
\label{PhaseCorrection}
\end{equation}
The proof for more than one turning point is straightforward and the scattering phase just add up. It would become clear in the following section that the scattering phase correction is crucially important in determine the energy spectrum.

\section{Modified EBK Quantization Rule}

The most powerful use of Van Vleck's formula is that it
leads to the EBK quantization rule in the semiclassical
limit.  One notices that, if we set $ x=x^{\prime}$ in the
propagator and integrate over all possible $x$, it results
in the quantum partition function $Z(T) = \sum_{n}
\exp[-iE_{n}T]$.  The energy levels can then be identified
as the singularities of $Z(\omega)$ which is the Fourier
transformation of the partition function.  Within stationary
phase approximation, it can be shown that the total
phase $\oint pdq - i\nu \pi/2$ (in the absence of sharp
boundaries) is quantized\cite{Kleinert95} and leads to the
EBK quantization rule,
\begin{equation}
\oint pdq = 2n\pi +\nu \frac{\pi}{2},  \label{EBKQuanta}
\end{equation}
where $\nu$ is the number of turning points along the
periodic orbit.  The usual WKB approximation is the special
case with two conjugate points $\nu =2$. The presence of the sharp boundaries changes the scattering phase at each turning point from $\pi/2$ to $\pi$ and leads to the modified EBK quantization rule. It is interesting to see that the scattering due to sharp confining potential modified the spectrum only through the scattering phase $\phi_s$.

Now we are ready to consider the confining potential in more general form
\begin{equation}
V_c(x) = \Theta(-x) [V_0 + V_1 |x|],
\end{equation}
where $V_0 \equiv k_0^2/2m$ is the potential height and $V_1 = k_1^3/2m$ is the slope of the confining potential. The scattering due to $V_c(x)$ can be solved exactly and the eigenstates are
\begin{equation}
|\psi(k)\rangle = |k\rangle + e^{-i\phi_s(k)} |-k\rangle.
\end{equation}
The scattering phase is apparently energy-dependent as shown in Figure 2. For the hard-wall potential $(k/k_0=0, k/k_1=0)$, the scattering phase is $\pi$, while for the smooth potential $(k/k_0\ll 1, k/k_1\gg 1)$ the phase becomes $\pi/2$ as in the WKB approximation. Following similar calculation in previous section, we arrive at the modified EBK quantization integral in Eq.~\ref{NewEBK}.

\begin{figure}
\centering
\includegraphics[width=0.8\linewidth]{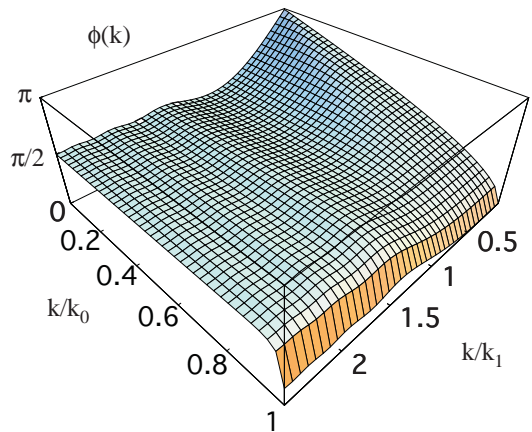}
\caption{Scattering phase for different potential height $V_0=k_0^2/2m$ and slop $V_1 = k_1^3/2m$.}
\end{figure}

We apply the modified EBK quantization rule to a finite potential well of length $L$ and with height $V_0 = k_0^2/2m$. After some algebra, the scattering phase is shown to be $\phi(k) = 2 \cos^{-1}[(k/k_0)^2-1]$. The quantized energy $E_n = k_n^2/2m$ satisfies
\begin{equation}
2 k_n L = 2 n \pi + \phi(k_n).
\end{equation}
Quite surprisingly, the spectrum obtained by the semiclassical approach is identical to the exact solution. This shows that the quantum interference effects arose from the sharp confining potential can be captured by the scattering phase rather well.

The modified EBK quantization rule can also be applied to
physical systems in higher dimensions.  Let us consider a
spherical or hemispherical qunatum dot.  We can either apply
the modified EBK formula directly to the true
three-dimensional trajectories\cite{Lin03} or apply the
formula after reducing the system to one dimension.  Here we
adapt the second approach.  After separation of variables,
the radial effective Hamiltonian of the three-dimensional
spherical (hemispherical) quantum dot becomes
one-dimensional with the effective potential
\begin{equation}
V =\left\{ \begin{array}{cl}
\frac{l(l+1)}{2mr^{2}}, &r<a\\
\infty, &r>a
\end{array}\right. ,
\end{equation}
where $l$ is the quantized angular momentum.  For the
spherical quantum dot, $l$ takes on all integer values,
while for the hemispherical dot, only odd integers are
allowed due to the flat boundary.

The classical trajectory of the electron is confined between
the hard-wall boundary at the surface and the centrifugal
potential near the origin.  Thus, there are one reflection
point $\phi_{s} = \pi$ and one conjugate point $\phi_{s}=\pi/2$.
Applying the modified EBK quantization rule, the approximate energy
satisfies the algebraic equation,
\begin{equation}
\sqrt{(a/r_{E})^{2}-1} 
-\sec^{-1}(a/r_{E}) 
= \frac{2\pi (n+\frac34)}{\sqrt{l(l+1)}},
\label{QDot}
\end{equation}
where $r_{E} = \sqrt{l(l+1)/(2mE)}$ is the conjugate point
and $a$ is the radius of the dot.  Instead of solving the
Schr\"odinger equation directly, the energy levels can be
determined easily by the algebraic equation in Eq.~\ref{QDot}.  In the
semiclassical limit, the conjugate point is close to the
origin, i.e., $a/r_{E} \gg 1$.  The approximate expression
can be further simplified,
\begin{equation}
E_{n,l} \approx \frac{\pi^{2}}{2ma^{2}} 
\left(n+\frac34+\frac{l'}{2}\right)^{2},
\label{EBKEnergy}
\end{equation}
where $l' = \sqrt{l(l+1)}$.

Notice that this problem can be solved exactly by the
spherical Bessel functions.  The hard-wall boundary requires
the wave function vanishes at the surface of the sphere,
$j_{l}(\sqrt{2mE}a)=0$, that leads to quantized energy
levels.  In the same limit $a/r_{E} \gg 1$, the spherical
Bessel function is approximated by the asymptotic expansion
that leads to
\begin{equation}
E^{ex}_{n,l} \approx \frac{\pi^{2}}{2ma^{2}} \left[ n+ \frac{l}{2}\right]^{2}.
\end{equation}
The above exact result does not seem to agree with
Eq.~(\ref{EBKEnergy}) at first glance.  However, if the
angular momentum is also semiclassical ($l \gg 1$), the last
term in Eq.~(\ref{EBKEnergy}) is $l'/2 \simeq l/2 + 1/4 $ up
to $O(1/l)$ corrections.  It is then clear that both give
the same result.  We emphasize again that the agreement is
only possible when the appropriate scattering phase is included.

Another way to obtain the modified EBK quantization rule for the $1/r$ potential is the conventional Langer's correction approach. Instead of including the appropriate scattering phase, one can obtain the same energy spectrum by modified the potential appropriately. While both approaches give the same spectrum, it is known that the wave functions calculation in scattering phase approximation is more accurate.\cite{Friedrich96,Friedrich99}

\begin{figure}
\centering\includegraphics[width=7cm]{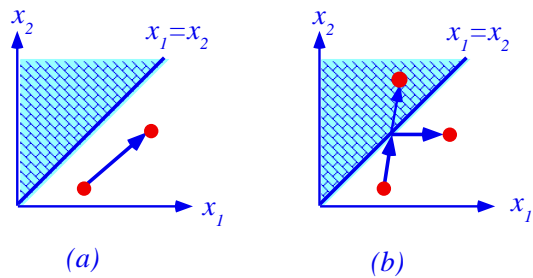}
\caption{Classical trajectories of two particles whose
quantum statistics is replaced by the equivalent hard wall
at $x_1=x_2$ in the configuration space.  In part (a), a
direct trajectory is shown and, in part (b), the shown
reflected trajectory is equivalent to exchanging two
particles which results in an extra Berry phase.}
\end{figure}

\section{Mirror Projection}

In the previous section, we treat the hard-wall boundary as
depletion of the Hilbert space.  An alternative way is to
view it as a discrete ${\cal Z}_2$ gauge symmetry of the
wave function
\begin{eqnarray}
\psi (x)=-\psi (-x).
\end{eqnarray}
The minus sign is chosen here to make the wave function
vanishes at $x=0$ so that the boundary condition $\psi
(0)=0$ is always satisfied.  Since the propagator can be
written down as the summation of eigenfunctions, $
G(x,x^{\prime };T)=\sum_n\psi _n^{}(x)\psi _n^{*}(x^{\prime
})\exp [-iE_nT],$ where $\psi _n(x)$ is the eigenfunction
with eigenenergy $E_n$.  The discrete gauge symmetry of the
wave function implies that the quantum propagator has the
symmetry
\begin{eqnarray}
G(x,x^{\prime };T)=-G(x,-x^{\prime };T).  \label{GaugeSymm}
\end{eqnarray}
Now choose both $x=r$ and $x^{\prime }=r^{\prime }$ to be
positive, the propagator can also be viewed as the wave
function $G(r,r^{\prime };T)=\psi _{r^{\prime }}(r,t)$ that
satisfies the Schr\"{o}dinger equation with a delta function
source at $(x,t)=(r^{\prime },0)$.  The propagator $
G_0(r,r^{\prime };T)$ without the hard-wall boundary
satisfies exactly the same differential equation except that
the boundary condition at $x=0$ is not met.  Notice that the
mirrored propagator $\overline{G}_0(r,r^{\prime };T) =
G_0(r,-r^{\prime };T)$ satisfies the Schr\"{o}dinger
equation without the source term since the delta function
$\delta (r+r^{\prime }) = 0$ for positive coordinates. 
Therefore, the propagator that satisfies the correct
boundary condition is constructed as
\begin{eqnarray}
G(x,x^{\prime };T)=
G_0(x,x^{\prime};T)-\overline{G}_0(x,x^{\prime };T).
\end{eqnarray}
The above result is equivalent to Eq.~(\ref{Propagator}). 
It is obvious that the discrete gauge symmetry in
Eq.~(\ref{GaugeSymm}) is satisfied.  This method is just the
familiar mirror charge trick in the classical
electromagnetism. 

Since we can solve the hard-wall boundary by discrete gauge
symmetry, we might as well go the other way around.  It is
possible to replace the quantum statistics between particles
by the hard-wall boundaries.  Let us consider the simplest
case -- two interacting particles with either bosonic or
fermionic statistics.  The discrete gauge redundancy is
\begin{equation}
\psi(x) = e^{i\phi} \psi(-x),
\end{equation}
where $x \equiv x_{1}-x_{2}$ is the relative displacement between two particles. 
The phase correction is $\phi =0$ for bosons and $\pi$ for fermions.  
The discrete gauge symmetry is removed by imposing a hard wall
$x_{1}=x_{2}$ in the configuration space, and a Berry phase
$\phi$ accumulates upon each reflection due to the hard
wall.

Classical trajectories are classified into two categories --
the direct path and the reflected one as shown in Figure 3. 
If we extend the reflected trajectory into the unphysical
regime inside the hard wall, as shown in Fig.  3(b), the
reflected trajectory is equivalent to an exchange between
two particles.  This approach would be useful when studying
few interacting quantum particles, e.g., two strongly
interacting bosons or fermions bouncing back and forth
between two hard walls.  In the semiclassical limit, we can
safely ignore the quantum statistics by solving all
classical trajectories inside a specific triangle in the
two-dimensional configuration space.

\section{Conclusions}

In this paper, we study the scattering phase of classical
trajectories due to sharp confining potentials. 
Inclusion of the energy-dependent scattering phase, 
the modified EBK quantization rule is derived.
We also relate the hard wall boundary
approach to the quantum statistics and the discrete gauge
symmetry.
Unlike the WKB approximation that is only applicable to
smooth potential profiles, the new quantization rule
provides us with an easy way to estimate the energy levels
in the presence of both smooth and sharp confinement
potentials.

We thank Darwin Chang for fruitful discussions, especially
on the mirror projection and the discrete gauge symmetry. 
This work was supported by the National Science Council of
Taiwan, R.O.C..

\appendix 
\section{Poisson Summation Formula}

Poisson summation formula provides a convenient way to
related two infinite summations together.  Let us consider a
physical system on a finite ring with length $L$ and lattice
constant $a$.  The total number of sites is $N=L/a$.  The
discrete version of the usual delta function is
\begin{equation}
\sum_{x=na}e^{ikx}=\left( \frac La\right) 
\sum_{G=2n\pi /a}\delta _{k,G},
\label{DeltaFunction}
\end{equation}
where $G$ is the reciprocal lattice vector.  Consider the
following summation,
\begin{equation}
\sum_nf(na)=\int \frac{dk}{2\pi }F(k)\sum_{x=na}e^{ikx},
\end{equation}
where $x_n=na$ and $F(k)$ is the Fourier transformation of
$f(x)$.  With the help of the identity in
Eq.(\ref{DeltaFunction}), the summation over coordinates is
turned into another summation over reciprocal momenta. 
Taking the thermodynamical limit $L\to \infty $, the
discrete delta functions are related to the continuous ones
by $L\delta _{k,G}=2\pi \delta (k-G)$.  Finally, we arrive
at the useful Poisson summation formula,
\begin{equation}
\sum_nf(na)=\frac 1a\sum_nF(\frac{2n\pi }a).
\end{equation}

\end{document}